\documentclass{pasa}%

\title[Millimetre-wave Site Characteristics at the Australia Telescope Compact Array]{Millimetre-wave Site Characteristics at the Australia Telescope Compact Array}
\author[Balthasar T. Indermuehle et al.]{Balthasar T. Indermuehle$^1$ \and Michael G. Burton$^2$\\
\affil{$^1$CSIRO Astronomy and Space Science, P O Box 76, Epping, NSW 1710, Australia, Email: balt.indermuehle@csiro.au}%
\affil{$^2$School of Physics, University of New South Wales, Sydney, NSW 2052, Australia}}%
\jid{PASA}
\doi{10.1017/pas.\the\year.xxx}
\jyear{\the\year}

\usepackage[authoryear]{natbib}
\bibpunct{(}{)}{;}{a}{}{,}
\usepackage{multirow}
\usepackage{tabularx}
\usepackage{booktabs} 
\usepackage{lscape}

\begin{document}%
\begin{abstract}
We present a millimetre-wave site characterisation for the Australia Telescope Compact Array (ATCA) based on nearly 9 years of data from a seeing monitor operating at this facility.  The seeing monitor, which measures the phase fluctuations in the signal from a geosynchronous satellite over a 230\,m baseline caused by water vapour fluctuations along their sight lines, provides an almost gapless record since 2005, with high time resolution.  We determine the root mean square (rms) of the path length variations as a function of time of day and season.  Under the assumption of the ``frozen screen'' hypothesis, we also determine the Kolmogorov  exponent, $\alpha$, for the turbulence and the phase screen speed.  From these, we determine the millimetre-wave seeing at $\lambda = 3.3$\,mm.  Based on the magnitude of the rms path length variations, we estimate the expected fraction of the available observing time when interferometry could be successfully conducted using the ATCA, as a function of observing frequency and antenna baseline, for the time of day and the season.  We also estimate the corresponding observing time fractions when using the water vapour radiometers (WVRs) installed on the ATCA in order to correct for the phase fluctuations occurring during the measurement of an astronomical source.
\end{abstract}
\begin{keywords}
 site testing -- Atmospheric effects -- Techniques: interferometric -- Instrumentation: interferometers
\end{keywords}
\maketitle%
\section{INTRODUCTION}
The Australia Telescope Compact Array (ATCA) is a radio interferometer located near the town of Narrabri in northwestern New South Wales, Australia at S30$^{\circ}$ 18'46" E149$^{\circ}$ 33'00". For over 10 years it was the only millimetre interferometer located in the southern hemisphere \citep{ATCAUserGuide:2011fk}. It was originally planned as a centimetre wavelength synthesis telescope with upgrades to the 12, 7 and 3 millimetre observing bands to be implemented at a later stage. This, along with some accounts of how the site selection came to agree on Narrabri, is described in detail in \cite{Frater:1992uq}. Following a proposal to fit out the array with millimetre wave receivers, a site testing programme was initiated to establish its feasibility. Longer term observations however were limited to opacity measurements of the atmosphere at 30 GHz while single snapshot observations were proposed at a wavelength of 3\,cm with the array antennae. This is described in an ATNF internal memo \citep{Hall:1992fk}. From these measurements the fraction of nights when the transmission is suitable for obervation at a given wavelength can be estimated, but they do not allow the phase stability for operating an interferometer as a function of baseline to be determined. In this paper we present a comprehensive analysis of the millimetre site characteristics at the ATCA based on 8.5 years of data and extending the analysis of \citet{Middelberg:2006vn}, which used the initial 1 year of data gathered with a seeing monitor that was installed in 2004.

\section{DATA DESCRIPTION}
The seeing monitor is a fixed baseline interferometer with a 230 m baseline and two receivers. They observe a transmitter beacon on a geosynchronous satellite. From 2004 until the satellite's decommissioning in September 2008, the signal used was the 30.48 GHz beacon on the Optus B3 satellite. The seeing monitor receiver hardware and software then was adapted to the new beacon frequency of 21.198 GHz on the Optus C1 satellite. The beacon signal is down converted and mixed to 50 kHz and sampled at 10 Hz with a phase locked loop (PLL) integration time of 100 ms. Individual phase measurements, however, are stored every 5 seconds only. The phases were corrected to their zenith values by multiplying by the cosine of the zenith angles (51$^{\circ}$ and 57$^{\circ}$ respectively). From this data, a zenith path difference $\Delta d$ is then formed for two successive 5\,s time intervals from $\lambda \Delta \Phi / 2 \pi$, where $\lambda$ is the wavelength of the beacon signal. A second order polynomial is fitted and used to eliminate satellite drift and thermal effects, as per the description in \cite{Middelberg:2006vn}. The standard deviation as well as the root mean square (rms) of this $\Delta d$ is computed for a 30 minute window. Because 30 minutes is much larger than the crossing time of a parcel of the atmosphere over the 230\,m baseline, using the standard deviation provides a robust measure of the magnitude of the phase fluctuations occurring over a small portion of the day. For all the parameters derived (see below) a 5$\sigma$ iterated sigma-clipping about the median value was also performed to remove deviant data. This resulted in the removal of no more than 2\% of the relevant values. The design of the seeing monitor, along with a site analysis spanning approximately one year of data has been described in further detail in \cite{Middelberg:2006vn} so we do not include it here.

We have examined 8.5 years worth of seeing monitor data spanning from April 2005 to October 2013. The site quality statistics we have derived include:
\begin{itemize}
\item the rms path length differences as a function of time of day and season,
\item the lag phase structure functions and Kolmogorov exponents, together with the phase screen speed,
\item the visibility efficiency as a function of baseline and frequency,
\item the millimetre-wave seeing at 3.3\,mm based on the path length fluctuations and Kolmogorov exponents. 
\end{itemize}
The Kolmogorov exponent determines whether the atmosphere above the ATCA can be treated as 2 dimensional ``frozen screen" (see below). The other parameters provide useful metrics for the fraction of time that conditions are sufficiently good to observe in a particular millimetre waveband. We conclude with an analysis outlining the improvement in observing time that could be achieved when using the ATCA water vapour radiometers (WVRs) described in \cite{Indermuehle:2013fk}, as a function of baseline, frequency, time of day and season.

\section{THEORY OF TURBULENCE}
\label{sec:theory}
Andrey Kolmogorov in 1941 developed the now widely accepted theory that small scale turbulent motions are statistically isotropic and thus do not have any preferential spatial direction \citep{Kolmogorov:1941uq}.  The large scale turbulent cells on the other hand are not isotropic because their behaviour is determined by the shape of the boundary.  Kolmogorov's hypothesis entertains the idea that a turbulent flow of velocity $V$ contains large vortices which over time decompose into smaller vortices until they have reached a scale where the remaining kinetic energy is turned into internal energy through viscous dissipation. Kolmogorov introduced the structure function in order to describe the fluctuations in the velocity field as a function of separation $r$:

\begin{equation}\label{sf}
D(r) = \int_{-\infty}^{\infty}dx(f(x)-f(x+r))^2
\end{equation}

which is equivalent to the mean square difference between two points of separation $r$:

\begin{equation}\label{sf2}
D(r) = \langle ( f(x) - f(x+r) )^2 \rangle 
\end{equation}

We model the layer containing the eddies of water vapour as a phase screen moving across the telescope aperture.  Taylor in his seminal paper on ``The Spectrum of Turbulence" \citep{Taylor:1938fk} found that for a given moving air mass with embedded eddies of varying properties such as temperature, humidity or water vapour, as long as the velocity of the air mass is very much greater than the turbulent velocity, the sequence of changes measured by a probe at a given point simply measures an unchanging pattern of eddies going by.  This means the eddies, as related to the probe, are not changing and seen as statically embedded (or ``frozen").  This is today referred to as the ``frozen screen" hypothesis (or Taylor hypothesis) and in our case applies to eddies of poorly mixed water vapour in the atmosphere. Assuming that the phase screen velocity, $v_s$, does not change significantly during the time period, $\tau$, over which a measurement is made, we can state that $\tau$ becomes equivalent to the baseline $b$ through the relationship

\begin{equation}\label{vs}
v_s=\frac{b}{\tau}
\end{equation}

where the lag time $\tau$ is the time it takes for the phase screen to pass over the baseline $b$.  We can then formulate the temporal phase structure function, or lag phase structure function (LPSF):

\begin{equation}\label{sf3}
D_\Phi(\tau) = \langle ( \phi(t) - \phi(t+\frac{b}{v_s}) )^2 \rangle 
\end{equation}

where $\phi$ is the phase, as modified by the water vapour.  By evaluating the LPSF using the seeing monitor data we can examine the scale of the turbulence, as well as the velocity of the phase screen $v_s$ through the following relationship:

\begin{equation}\label{sf4}
D_\Phi(\tau) = D_\Phi(b) \; | \; b=v_s\tau
\end{equation}

where $D$ is the phase structure function of the turbulence.  The square root of this corresponds to the rms phase variation:

\begin{equation}\label{sf6}
\Phi_{rms}(\tau) \equiv \sqrt{D_\Phi(b)}
\end{equation}

This can be formulated in a general way to determine a baseline and wavelength dependent term $K$ as shown by \cite{1990ursi.symp...11C} and \cite{1999RaSc...34..817C}:

\begin{equation}\label{sf5}
\Phi_{rms}(b)=K(\frac{\lambda_0}{\lambda})(\frac{b}{b_0})^{\alpha}
\end{equation}

where $\lambda$ is the observing wavelength, $\lambda_0$ that of the seeing monitor, $b$ is the antennae baseline length and $b_0$ the separation of the seeing monitors.  $\alpha$ is the Kolmogorov exponent and $K$ is measured in radians.  With $\lambda = \lambda_0$ and $b = b_0$ then $\Phi_{0,rms} = K$ is the rms phase fluctuation measured with the seeing monitor itself. According to Kolmogorov's theory of turbulence, the phase noise \textit{vs.}\  baseline relationship should follow equation \ref{sf5} with $\alpha=\frac{1}{3}$ for turbulence with baselines longer than the width of the tubulent layer (i.e.\ $b > h$).  This is referred to as 2D turbulence, because the phase screen assumes the statistical properties of a 2 dimensional, thin screen.


The other case is where the baseline is shorter than the width of the turbulent layer ($b < h$) and has 
a Kolmogorov exponent of
$\alpha=\frac{5}{6}$.  The phase screen then exhibits 3D behaviour and
is called the thick screen.


There is also an outer scale, $L_0$, beyond which the rms phase variation should no longer increase with baseline length.  This corresponds to $\alpha = 0$. We cannot determine the outer scale however because we are not measuring multiple baselines at a single point in time. 

The phase variations averaged over a given integration time cause coherence loss in the measured visibility.  For a given visibility $V=V_0e^{i\phi}$ the effect on the measured amplitude due to phase noise in a given averaging time is:

\begin{equation}\label{vis}
\langle V \rangle = V_0 \times \langle e^{i\phi} \rangle = V_0 \times e^{-\phi_{rms}^2/2}
\end{equation}

This assumes Gaussian random phase fluctuations of $\phi_{rms}$ over the averaging time. This is discussed in more detail in \cite{1999RaSc...34..817C}.

\section{THE RMS PATH LENGTH DIFFERENCE}
\label{sec:rms}
The seeing monitor provides only two output parameters: (zenith) phase and time. Taking these we can build the phase differences between adjoining measurements. The phase difference root mean square (rms) and its standard deviation can then be analysed. Because the (zenith) path differences fluctuate around a mean near zero, there is minimal difference between these measures. We use rms as the preferred metric. The phase differences can easily be converted to path differences using the frequency of the observed signal, $d= \lambda \Delta \Phi / 2 \pi$. 

\begin{figure*}[h]
\begin{center}
\includegraphics[width=2\columnwidth]{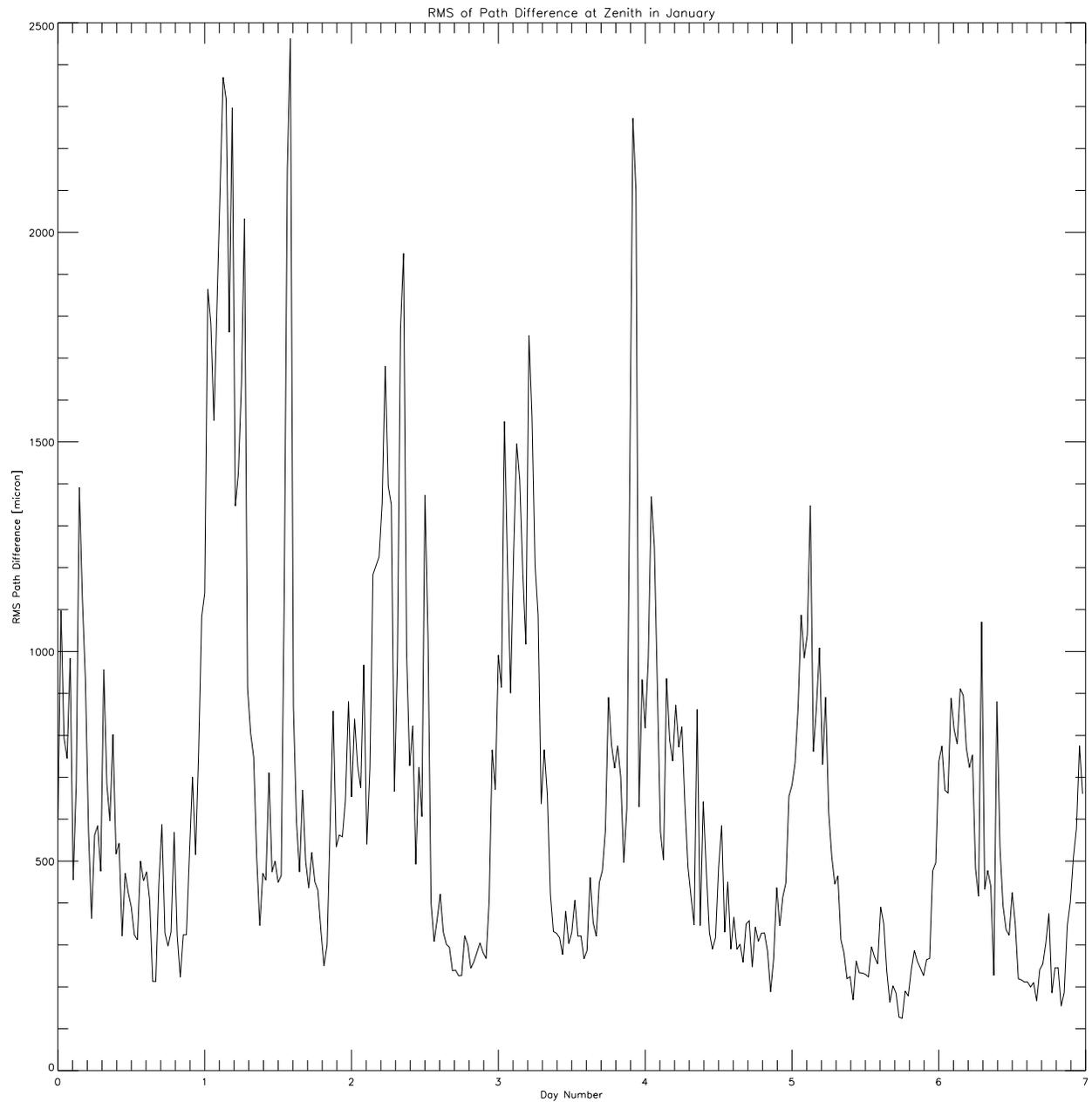}
\caption{The rms path difference for a week's worth of data in January. The range of path fluctutation varies by about an order of magnitude, from $\sim 250\, \mu$m at night to $\sim 2,500\,\mu$m during the day time.}
\label{figure1}
\end{center}
\end{figure*}

\begin{figure*}[h]
\begin{center}
\includegraphics[width=2\columnwidth]{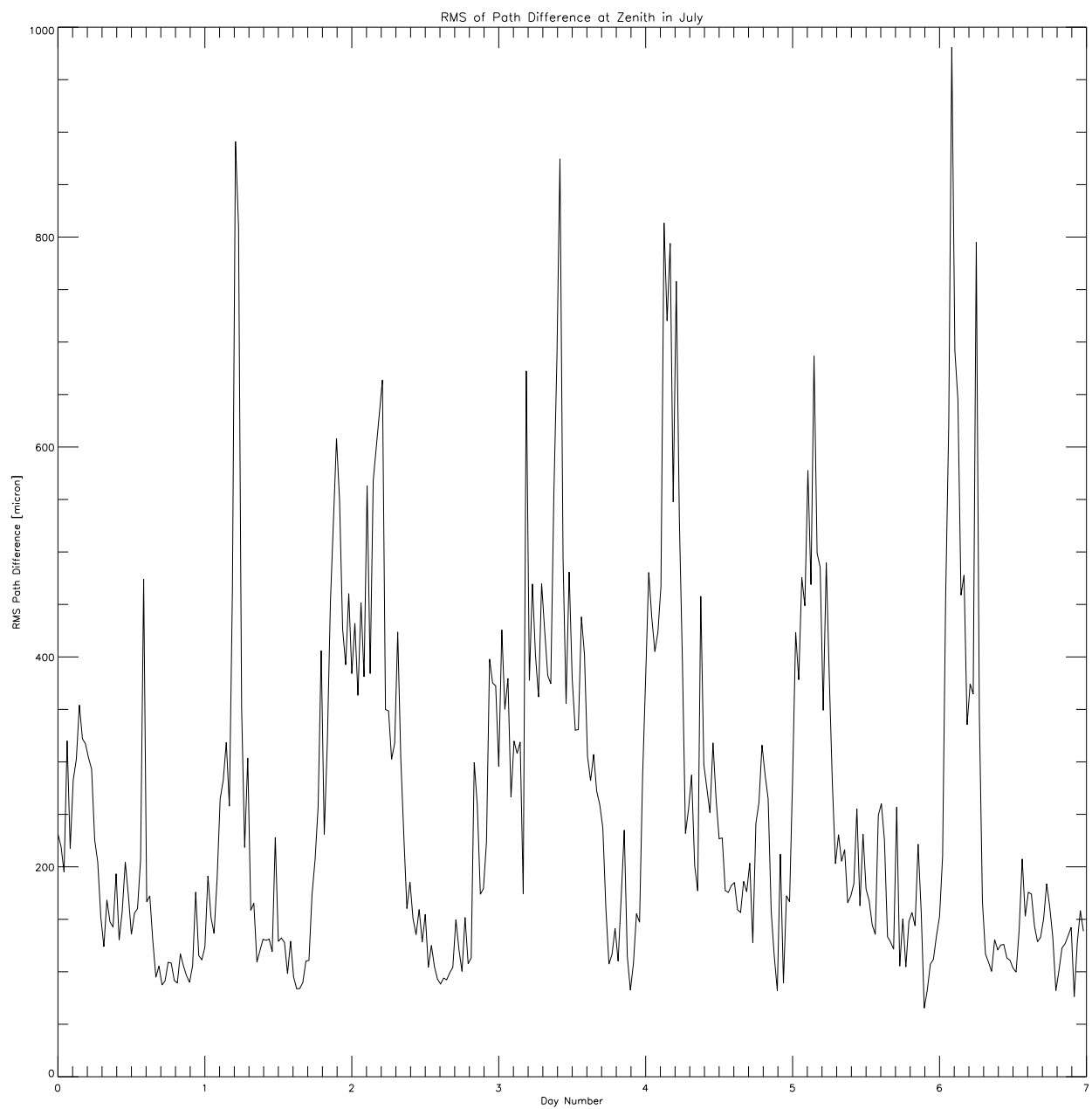}
\caption{The rms path difference for a week's worth of data in July. The fluctuations are similar in magnitude to the summer data in Figure \ref{figure1}, but their overall level is much lower, varying between 100 -- 950 $\mu$m. }
\label{figure2}
\end{center}
\end{figure*}

Figures \ref{figure1} and \ref{figure2} give examples of a week of path variations in January and July. The diurnal range of path fluctutation varies by about an order of magnitude in both the summer and winter data, but the amplitudes are higher in summer by a factor of about two. In the summer, they vary from about 250\,$\mu$m at night to 2,500\,$\mu$m during the day while in the winter, 100 -- 950\,$\mu$m of diurnal variation is more common. 

\begin{figure*}[h]
\begin{center}
\includegraphics[width=2\columnwidth]{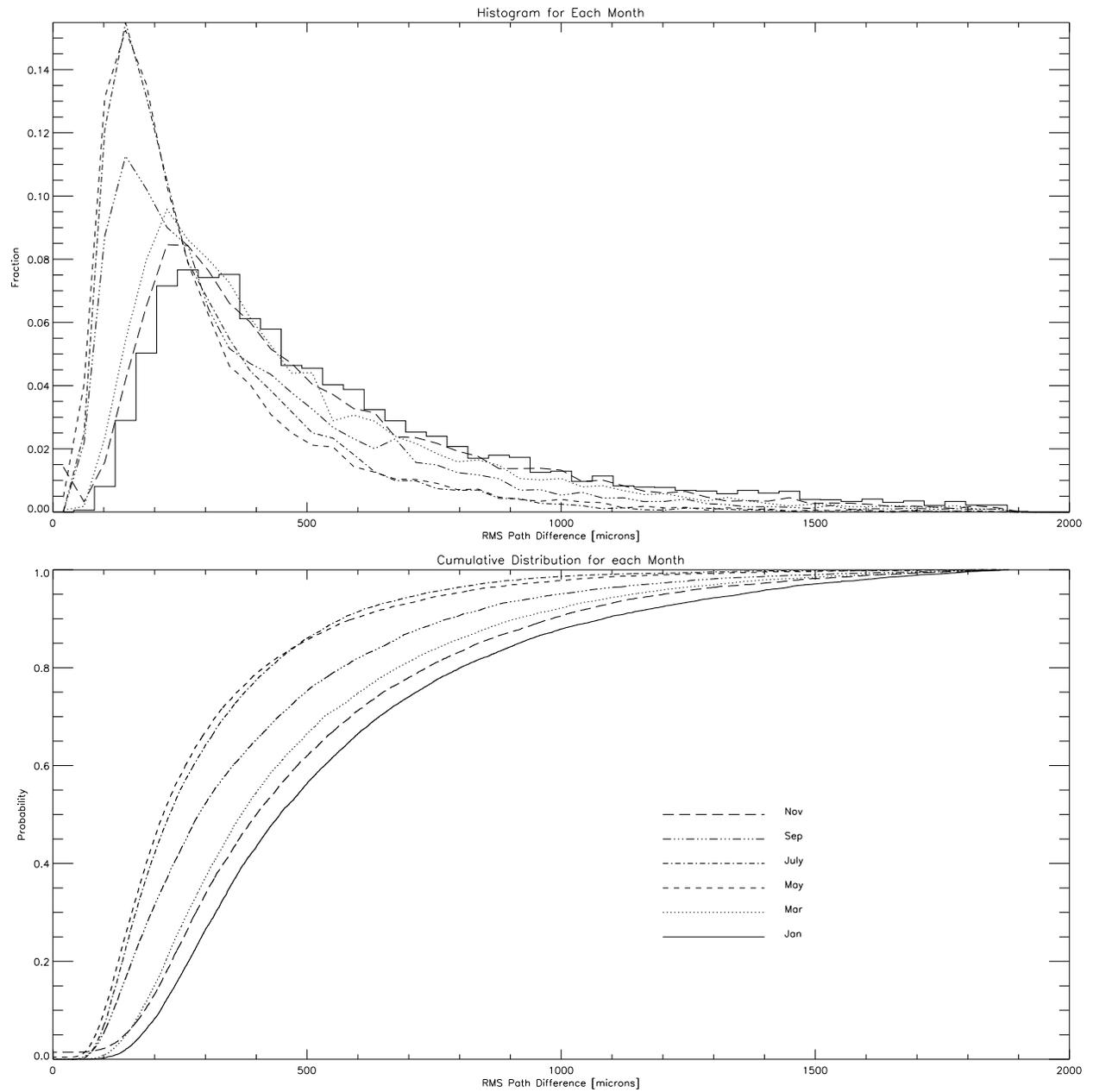}
\caption{Histogram and cumulative distribution of the rms path differences in $\mu$m. The differences between summer and winter months are clear, as discussed in \S\ref{sec:rms}.}
\label{figure3}
\end{center}
\end{figure*}

As can be seen from Figure \ref{figure3}, which shows the histogram and cumulative distribution of the rms path length differences through the year, the variations between the summer and winter months are clear. In May and July, for 80\% of the time the rms path difference is less than 400\,$\mu$m, whereas in January, for the same fraction of time, the rms path length is twice as high, i.e.\ about 800\,$\mu$m. The histogram features a marked peak for both the May and July curves, indicating a prolonged period of time (lasting from May to August) where rms path difference conditions of just 150\,$\mu$m or less are experienced for one-third of the time. Examination of Table \ref{table1}, where these values are tabulated, confirms that all the winter months feature rms values close to their minimum value.

\begin{figure*}[h]
\begin{center}
\includegraphics[width=2\columnwidth]{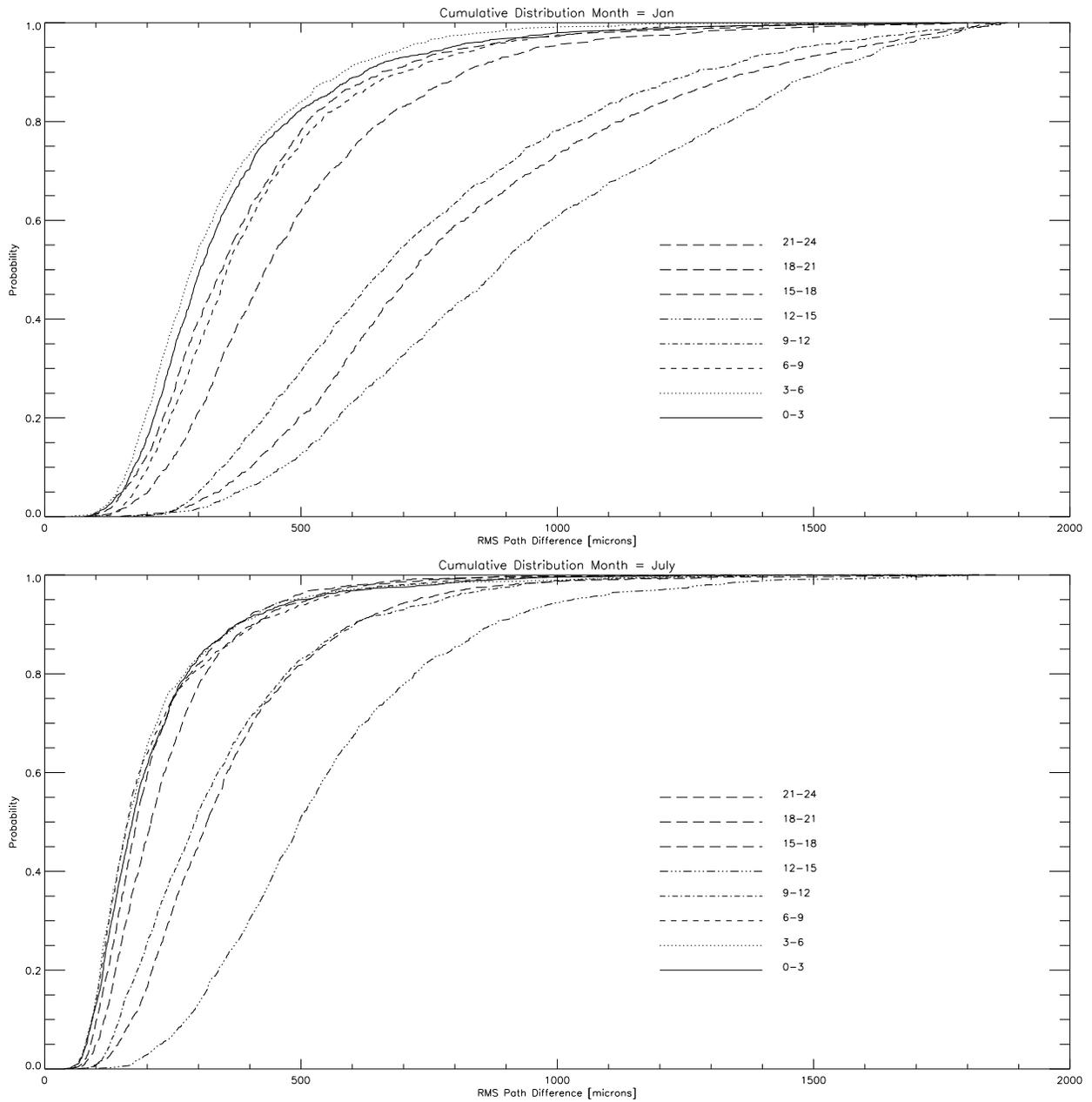}
\caption{Cumulative distributions of the zenith rms path differences in $\mu$m for time of day (in 3h bands) in summer (January, top) and winter (July, bottom). Summer nights have similar values to winter days.}
\label{figure4}
\end{center}
\end{figure*}

By further breaking this data down into time of day (i.e.\ 3\,h time bands) statistics, as shown in Figure \ref{figure4}, it can be seen that the worst conditions in winter, ocurring between 12 -- 15\,h LT (with about 750\,$\mu$m rms path difference), are similar to the average conditions in summer. However, while these are considerably better than conditions in the worst periods experienced during summer (12\,h -- 21\,h LT), they also are worse than the best periods in summer, which last from 21 -- 09 LT, i.e.\ from late night to mid-morning. The slope of the steeply rising portion of the cumulative distribution in Fig.~\ref{figure4} is steeper for the winter data and the difference between daytime and nighttime conditions is half of that in summer, indicating that the good conditions in winter last for a longer period of time; for roughly 15\,h per day (between 21 -- 09 LT), there is an 80\% probability of the rms path noise being less than 250\,$\mu$m. In comparison, during the best times in summer, lasting for 12\,h per day, there is an 80\% probablility of the path noise being less than 500\,$\mu$m (twice the typical winter-time rms path difference).

The distribution of the rms path differences, split into quartiles for each month and 3 hour time band, are listed in Table \ref{table1}, together with their maximum and minimum values within each month/time band. The maximum and minimum values encountered in these quartiles over the entire year are listed in Table \ref{table2}. As an illustrative example of the use of these Tables, in April at 12 noon, for 50\% of the time the path differences will be less than 728\,$\mu$m. The lowest value for the third quartile path difference occurs in May between midnight and 3\,am, and is 224\,$\mu$m.

A direct comparison of our results with the analysis of \cite{Middelberg:2006vn} is not readily possible as these authors convert the rms path differences they determine to a 1\,km baseline, making use of the Kolmogorov exponent they find for each path difference (i.e.\ scaling using eqn.~\ref{sf5}).  We instead provide the statistics in Tables~\ref{table1} and \ref{table2} for the 230\,m seeing monitor baselines as these can be used for direct comparison with the values provided to the observer by the ATCA control system, to assist in deciding upon the appropriate project to pursue at a particular time. We discuss how this information can be used in \S\ref{sec:efficiency}.  We have, however, compared our results to \cite{Middelberg:2006vn} by converting the rms path differences we measure to a 1\,km baseline using the local value for the Kolmogorov exponent.  The distributions we see in June and November, as per those shown by \cite{Middelberg:2006vn} in their Fig.~5, are broadly similar.  For instance, in June for roughly 80\% of the time the 1\,km path differences are better than 600$\mu$m for 15 hours per day in both our studies. The corresponding differerence is 1,000\,$\mu$m in November.  However, for the lowest quality periods, we find there is significantly more time with large rms differences over the 9 {\it vs}.\ 1 year range of the two data sets.  For instance, in November between 12--15 hours, we find that for roughly half the time the path differences are greater than 1,500\,$\mu$m, in comparison to $< 10$\% of the time by \cite{Middelberg:2006vn}.  Year to year variations can be significant, and might account for this difference. 

\section{DERIVED PARAMETERS}
We can derive further information from the rms path length differences by analysing the behaviour in the time domain, applying the theory outlined in \S\ref{sec:theory}.  This yields the Kolmogorov exponent ($\alpha$), corner time ($t_c$), saturation path length ($p_{sat}$) and phase screen speed ($v_s$), as we discuss below.

\subsection{The Kolmogorov exponent}
\label{sec51}
\begin{figure*}[h]
\begin{center}
\includegraphics[width=2\columnwidth]{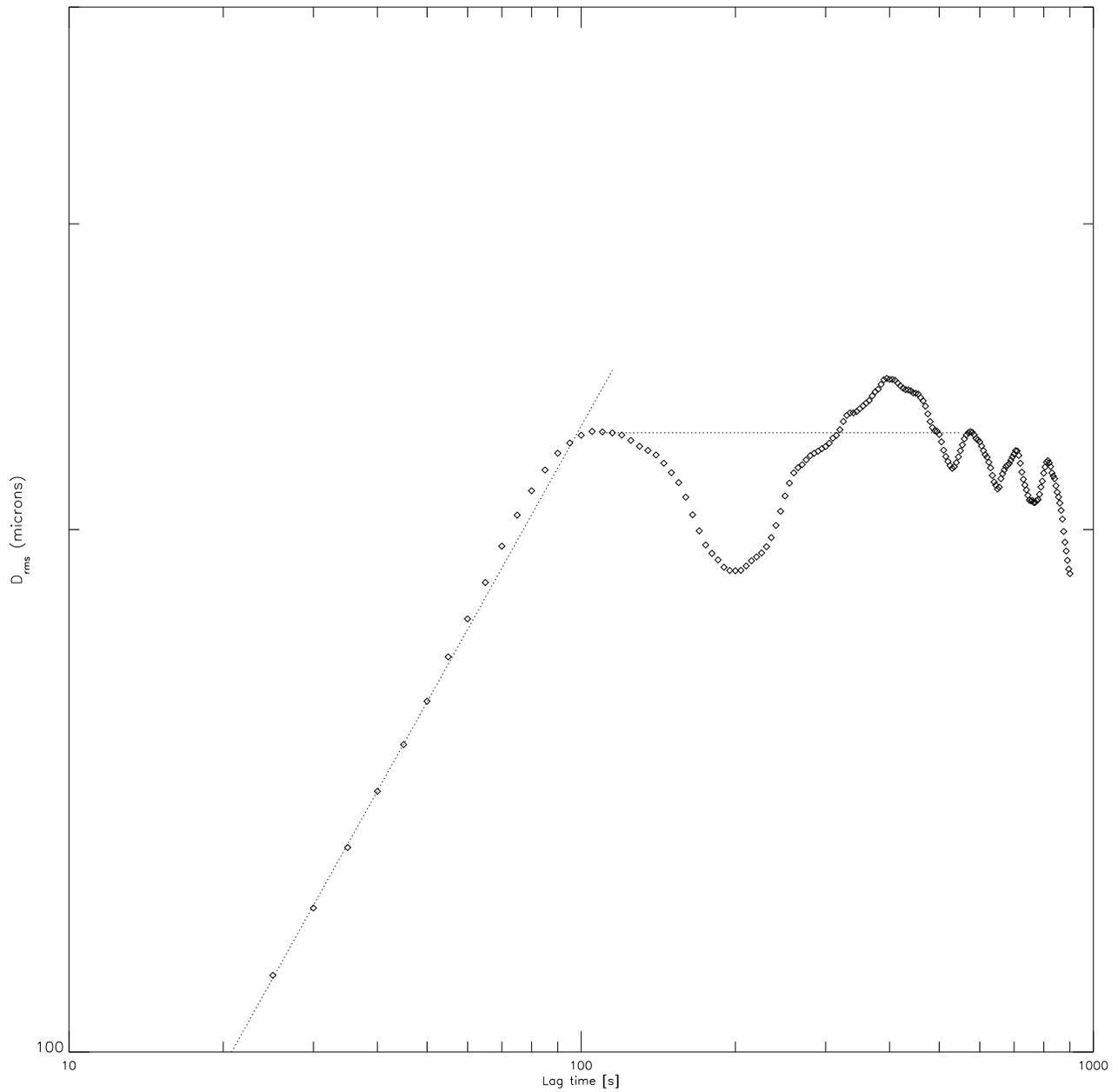}
\caption{An example lag phase structure function on June 16 2006, plotting lag time (s) against the rms path difference ($\mu$m). The horizontal line shows where the saturation path length was determined as the structure function's first peak and the slope of the fit to the rising portion yields the Kolmogorov exponent. In this example, the rms path length was found to be 227 $\mu$m at the corner time $t_c = 110$s, yielding a phase screen speed of 2.1 m/s and Kolmogorov exponent $\alpha=0.53$.}
\label{figure5}
\end{center}
\end{figure*}

Figure \ref{figure5} shows an example of a lag phase plot used to extract the lag phase structure function parameters. This illustrates the behaviour of the fluctuating phase screen as it moves across the baseline. As long as the structure of the air mass is not self similar, the root mean square (rms) of the path length variation in the measurement span increases to a maximum after which subsequent data result in a reduction in the rms because a self-similar scale size has been reached. This behaviour continues until the scale size of the turbulent layers in the atmosphere have become too extended and the data have become statistically independent, thus marking the point where the outer scale length for the turbulence has been reached.

The parameters that can be derived from the lag phase structure function include the corner time $t_c$, which is the time that elapses until the phase noise has peaked for the first time as well as the saturation path length noise $p_{rms,sat}$.  We determine the corner time as the time when two or more successive path length noise points are smaller than the preceding points. The saturation path noise $p_{rms}$ at $t_c$ can then be compared to the standard deviation of the path difference $\sigma_{pd}$ and because the data can be approximated by a series of sinusoids fluctuating about a zero point, it should satisfy $p_{rms} \approx \sigma_{pd} \times \sqrt{2}.$\footnote{The rms of any sinusoid is equal to its amplitude / $\sqrt{2}$.}  We then use the corner time $t_c$ and the baseline length $b$ to determine the velocity of the phase screen $v_s$ according to the relationship shown in equation \ref{vs}.  The gradient fit in log space to the phase noise before the corner time has been reached determines the Kolmogorov exponent $\alpha$.

\begin{figure*}[h]
\begin{center}
\includegraphics[width=2\columnwidth]{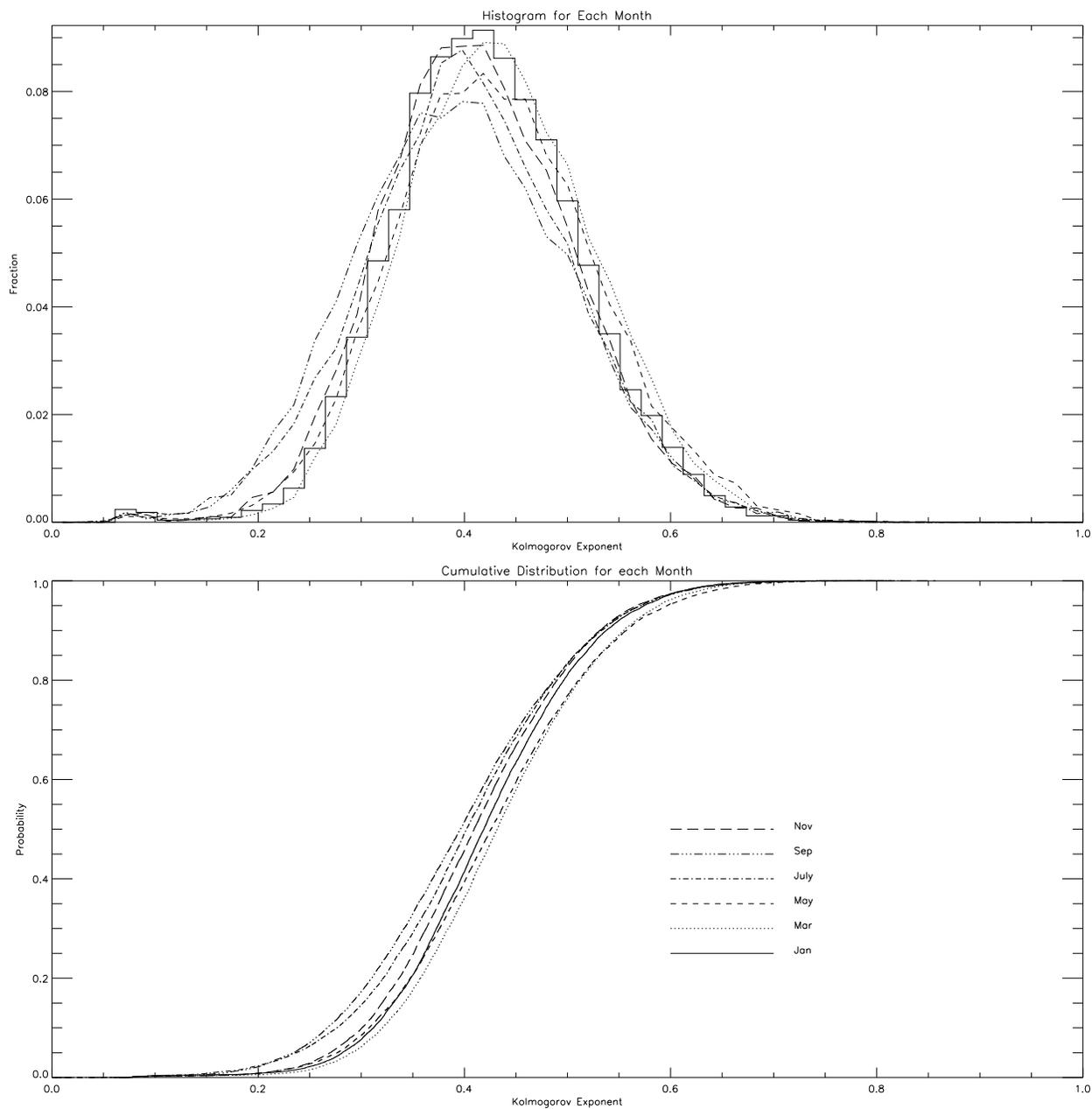}
\caption{Histogram and cumulative distribution of the Kolmogorov exponent $\alpha$ as a function of the month. Seasonal variations are seen to be small. For clarity, only every other month is shown. }
\label{figure6}
\end{center}
\end{figure*}

\begin{figure*}[h]
\begin{center}
\includegraphics[width=2\columnwidth]{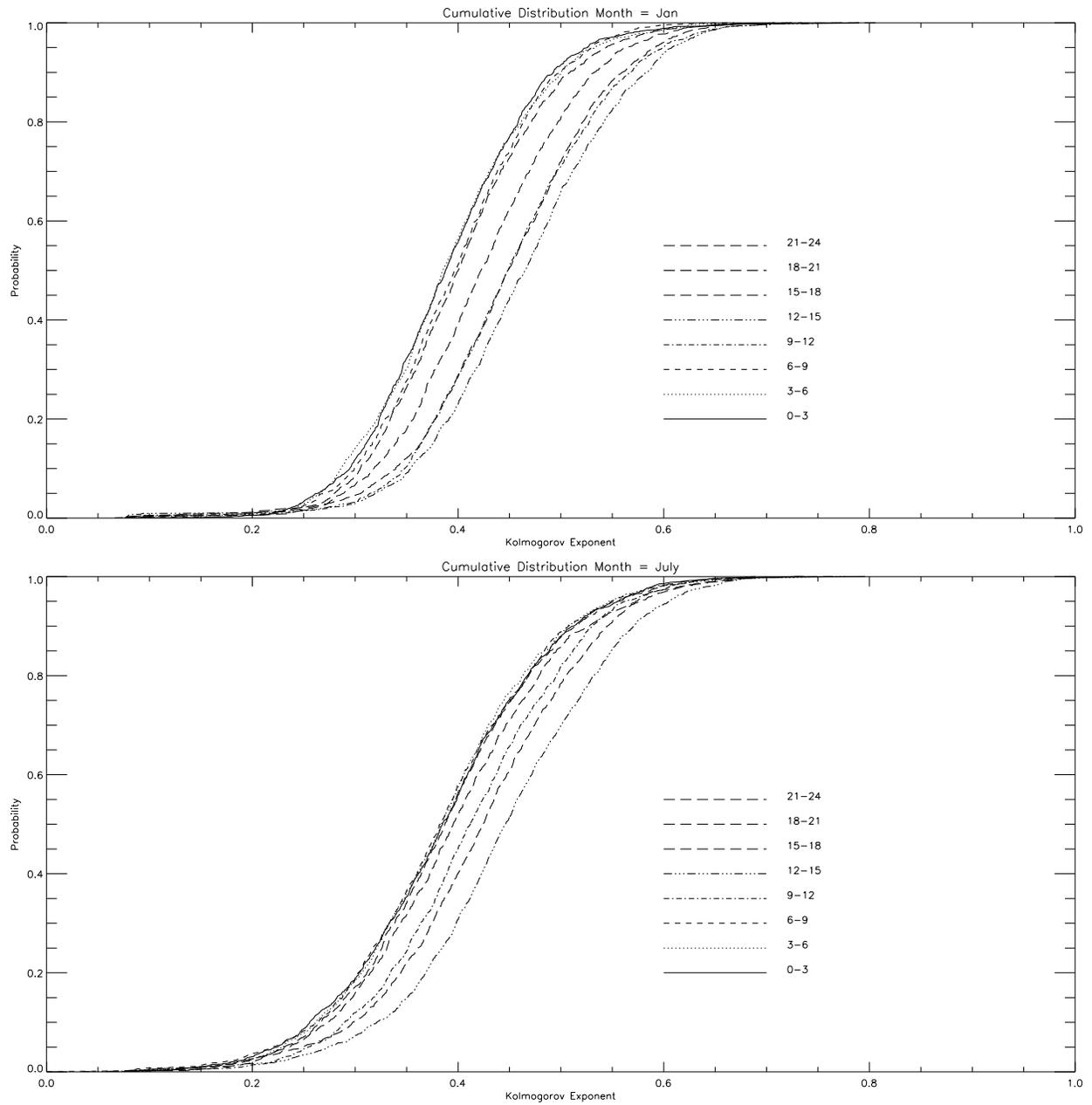}
\caption{Cumulative distribution of the Kolmogorov exponent $\alpha$ for the months of January (top) and July (bottom) shown in three hour time bands. Diurnal variations are relatively small, but are larger than the seasonal variations shown in Fig.~\ref{figure6}. }
\label{figure7}
\end{center}
\end{figure*}

In Figure \ref{figure6} we show the histogram and cumulative distribution of the Kolmogorov exponents for the complete date range from 2005 -- 2013.  The distributions for the time of day (in 3 hour bands) for July (i.e.\ winter)  and January (i.e.\ summer) are shown in Figure \ref{figure7}. Neither the diurnal nor the seasonal variations are large. As can be seen from Table \ref{table4}, the median value is 0.41, and the median ranges between 0.40--0.43 over the months of the year. The time of day dependence has a preference towards a larger $\alpha$ during day time hours from 09:00 to 18:00 local time. The slope of the January cumulative distribution in Figure \ref{figure7} is steeper than in July with their peaks being approximately the same, $\alpha=0.65$. At the lower end however, the winter time Kolmogorov exponent extends to smaller values by about $\Delta\alpha\approx0.1$. Variations in the Kologorov exponent are clearly dominated by the time of day rather than the season. 

From these data it is evident that for the great majority of time the site experiences 2D turbulence since $\alpha$ is close to the value of $1/3$ (see \S\ref{sec:theory}).  This means that the vertical extent of the turbulent layer is thinner than the length of the baseline (i.e.\ \textless\ 230 m). $\alpha$ is never as large as 5/6 as would be required for 3D turbulence. For 50\% of the entire time $\alpha = 0.41$ with the lowest 50\% quartile located in August with $\alpha=0.40$ and the highest in March with $\alpha=0.43$, a relatively small annual variation.
The largest values for $\alpha$ found in summer, occurring from 09--18 h, are consistent with there being an agitated troposphere caused by the Sun's thermal heating at these times. Conversely, the smaller values in winter suggest a thinner phase screen than in summer, on average, presumably due to there being lesser thermal heating in winter.

\subsection{Phase Screen Speed}
\begin{figure*}[h]
\begin{center}
\includegraphics[width=2\columnwidth]{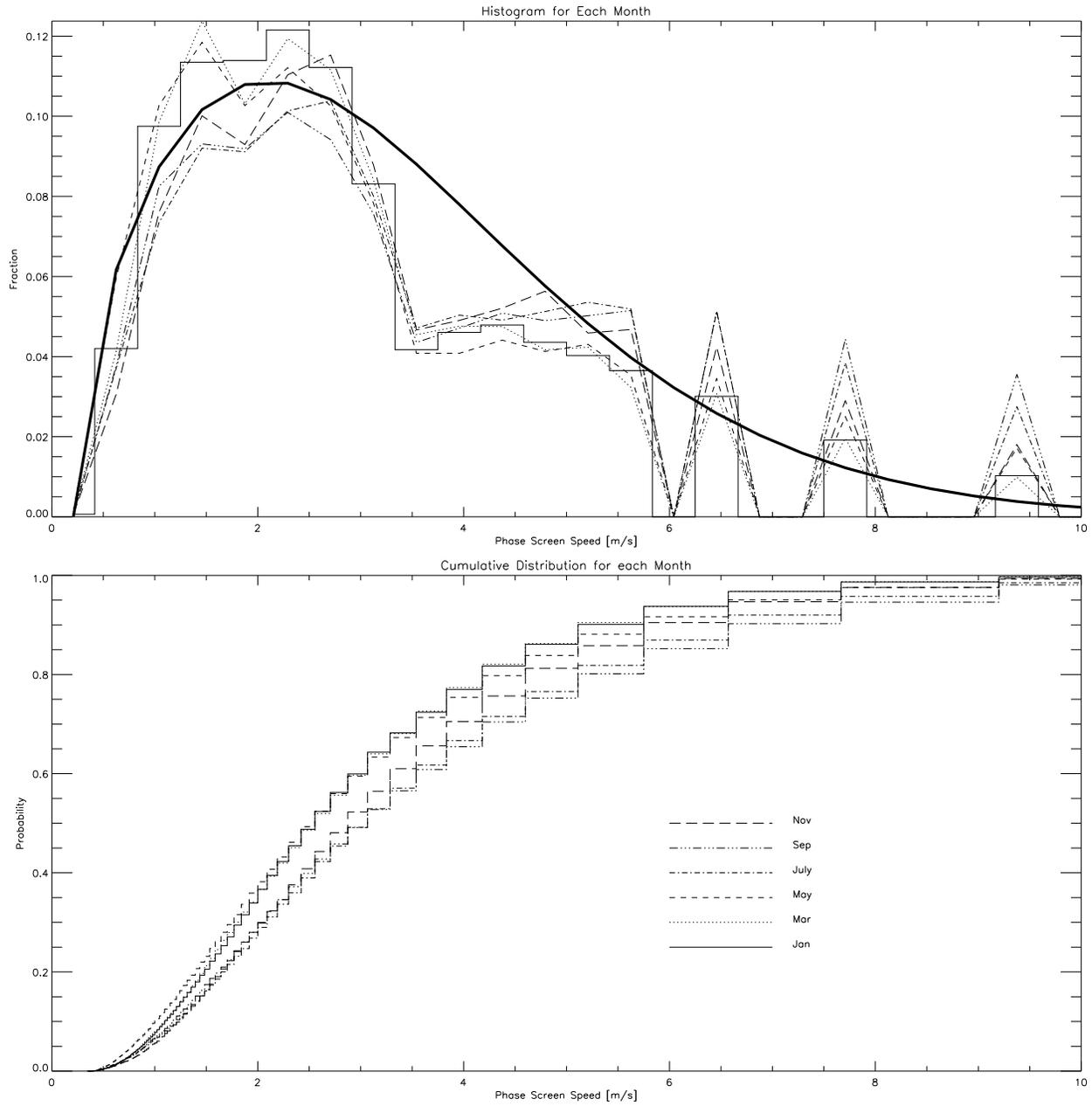}
\caption{Histogram and cumulative distribution of the phase screen speed distribution as a function of month (for clarity only every other month is shown). Note that the quantisation effect arises out of the time resolution of 5 seconds. The thick line overlay in the histogram plot shows the Weibull distribution as shown in Equation \ref{weibull}, with location parameter $\gamma\sim3.5$, shape parameter of $\beta\sim1.6$ and scale parameter $\eta \sim 2.0$. }
\label{figure8}
\end{center}
\end{figure*}

From the lag phase structure function the corner time is obtained, then, by using equation \ref{vs}, the phase screen speed. In Figure \ref{figure8}, we show the histogram and cumulative distribution for the phase screen speed on a monthly basis for the entire 9 years of data. The histogram conforms to a classical Weibull distribution,


\begin{equation}\label{weibull}
F(x,\beta,\gamma) = \frac{\beta}{\eta} \left( {\frac{x-\gamma}{\eta}} \right) ^{\beta-1} e^{-(\frac{x-\gamma}{\eta})^\beta}
\end{equation}

In this case a location parameter $\gamma\sim3.5$, scale parameter $\eta \sim 2.0$ and shape parameter of $\beta\sim1.6$ provide a good fit.  This later value is between that for an exponential ($\beta=1$) and a Rayleigh ($\beta=2$) distribution. Wind speed data throughout Meteorological literature is often seen to follow a Weibull distribution \citep{D.-N.-Prabhakar-Murthy:2004fk}.

\begin{figure*}[h]
\begin{center}
\includegraphics[width=2\columnwidth]{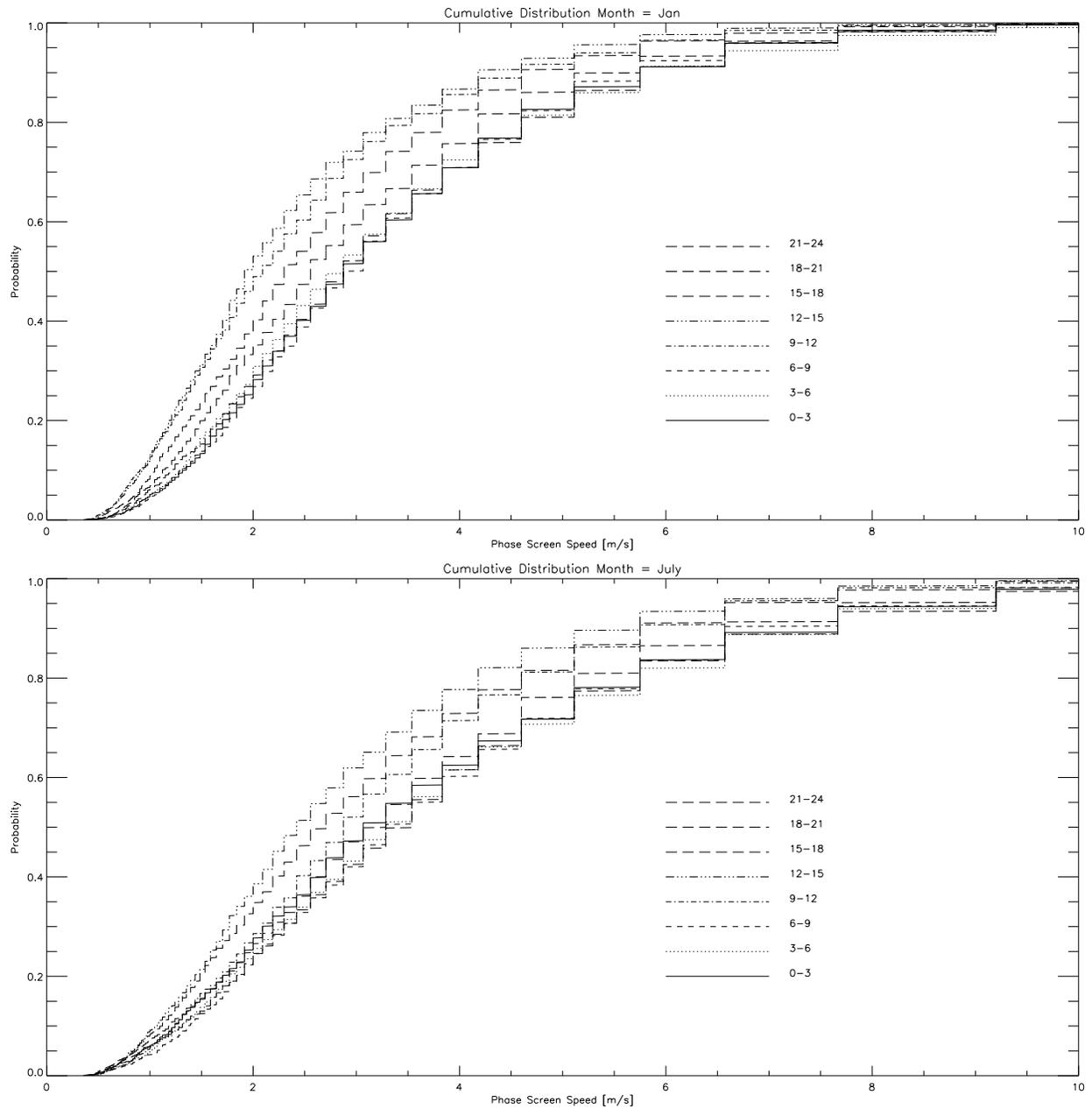}
\caption{Cumulative distributions for the phase screen speed as a function of time of day (in three hour bands) in summer (January, top) and winter (July, bottom).}
\label{figure9}
\end{center}
\end{figure*}

Of note is the quantisation effect due to the time resolution (5s) as well as the minimal seasonal variation in the phase screen speed. While in winter there are slightly higher wind speeds, the difference with summer is small, of order 0.5 m/s.  In Figure \ref{figure9} the cumulative distributions for the phase screen speed are shown for 3-hour time bands in January and in July.  A complementary picture to Figure \ref{figure4} is seen. The lowest wind speeds occur during the same time span, between midnight and 9 am, when the rms path variations are also lowest. This is consistent with the flow being more laminar and thus less turbulent when it is associated with lower wind speeds.

\subsection{Seeing limitations}
\label{sec:seeing}
Another important effect caused by tropospheric phase fluctuations is the limitation it imposes on the ability of the telescope to spatially resolve structure. Analogous to the seeing limitations caused by microthermal fluctuations in the optical regime, in the radio regime the water vapour induced phase delays cause a positional variation in the source observed because interferometric phase corresponds to the measurement of a point source \citep{Perley:1998kx}.  

Following \cite{1999RaSc...34..817C}, the half-power point in the visibility (eqn.~\ref{vis}) occurs when $\Phi_{rms} = 1.2$ radians.  From the seeing monitor measurements $\Phi_{0,rms} = p_{rms}/\lambda_0 \times 2 \pi$ where $p_{rms}$ are the rms path differences.  Then, using eqn.~\ref{sf5} and substituting in eqn.~\ref{vis} we can define a half-power baseline length (HPBL), $b_{1/2}$, by
\begin{equation}\label{bhalf}
b_{1/2} =  (\frac{1.2 \lambda}{K \lambda_0})^{1/\alpha}   
\end{equation}
with $K = \Phi_{0,rms}$.


The seeing is then given by

\begin{equation}\label{sf10}
\theta = \frac{\lambda}{b_{1/2}}
\end{equation}

For example, on June 16 2006, the phase structure function parameters were determined, as shown in Figure \ref{figure5}. The rms path difference was 227\,$\mu$m, the corner time 110\,s and therefore the phase screen speed was 2.1\,m/s and Kolmogorov exponent $\alpha=0.53$.  From this we obtain $K = 0.095$ radians ($5.4^{\circ}$).   For observations to be conducted at 3.3\,mm (90\,GHz), with $\lambda_0 = 15$\,mm and $b_0 = 230$\,m,  then the half power baseline is 1.6\,km and the corresponding millimetre seeing is 0.43 arcsec.

In Table \ref{table3}, the limiting seeing values for an observing wavelength $\lambda=$ 3.3mm are listed based on the calculated half power baselines for the median Kolmogorov exponents and path difference rms for each hour of the day during the best (June) and worst (December) periods of the year. The best seeing conditions are of order 0.3 arcsec and  occur in June, between midnight and 6 am. The worst conditions occur during December midday, and are about 2.5 arcsec.

\subsection{Correlations between Derived Parameters}
\begin{figure*}[h]
\begin{center}
\includegraphics[width=2\columnwidth]{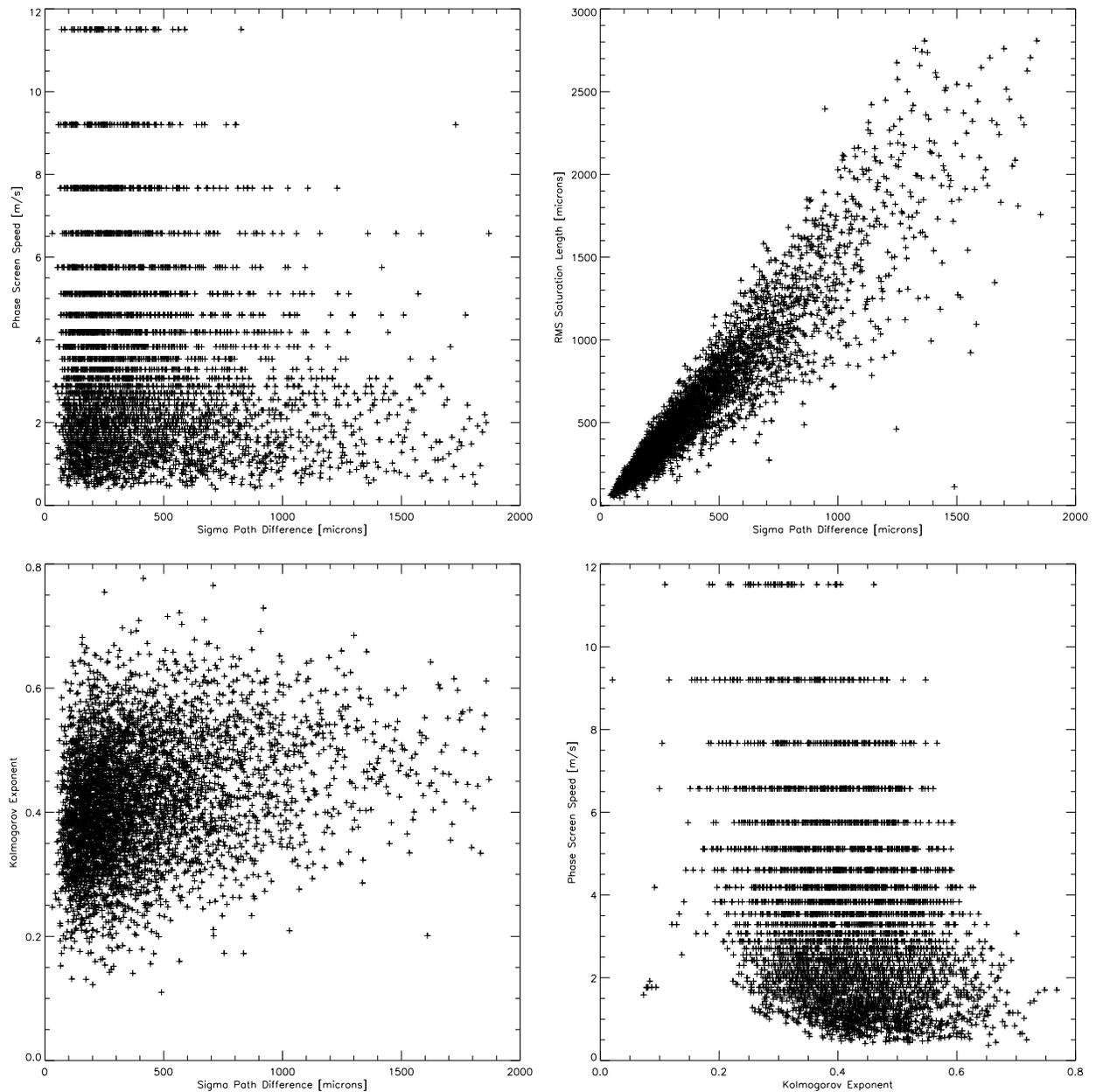}
\caption{Correlations between the derived values for the rms path difference and phase screen speed (top-left), rms path difference and saturation path length (top-right), rms path difference and Kolmogorov exponent (bottom-left) and the Kolmogorov exponent and phase screen speed (bottom-right). The corresponding correlation coefficients are -0.12, 0.93, 0.25 and -0.28 respectively. Note that the quantisation in two of the plots arises from the limited values possible for the phase screen speed (see text). For clarity, only every 30th point is plotted.}
\label{figure10}
\end{center}
\end{figure*}

In Figure \ref{figure10} we present scatter plots to search for correlations between the parameters derived from the seeing monitor data. As is evident from these, only the saturation path length and the rms path difference correlate strongly, with the slope to a linear fit being 1.44 $\pm$ 0.01, i.e. close to $\sqrt{2}$, as anticipated (see \ref{sec51}). There is a weak positive correlation between the Kolmogorov exponent and the rms path difference and a weak anti-correlation between it and phase screen speed. While the Kolmogorov exponent allows us to infer that the turbulence generally approximates 2D behaviour, the phase screen speed is not therefore useful in inferring the millimetre-site conditions. 

Only every 30th point is shown in Fig. \ref{figure10} for clarity. The alternative (and as shown by \cite{Middelberg:2006vn}) is to average the data in bins. This, however, results in a relatively smooth line which makes the correlation between the parameters appear to be better than it actually is.

\subsection{Available Observing Time in the Millimetre Wavebands}
\label{sec:efficiency}
\begin{figure*}[h]
\begin{center}
\includegraphics[width=2\columnwidth]{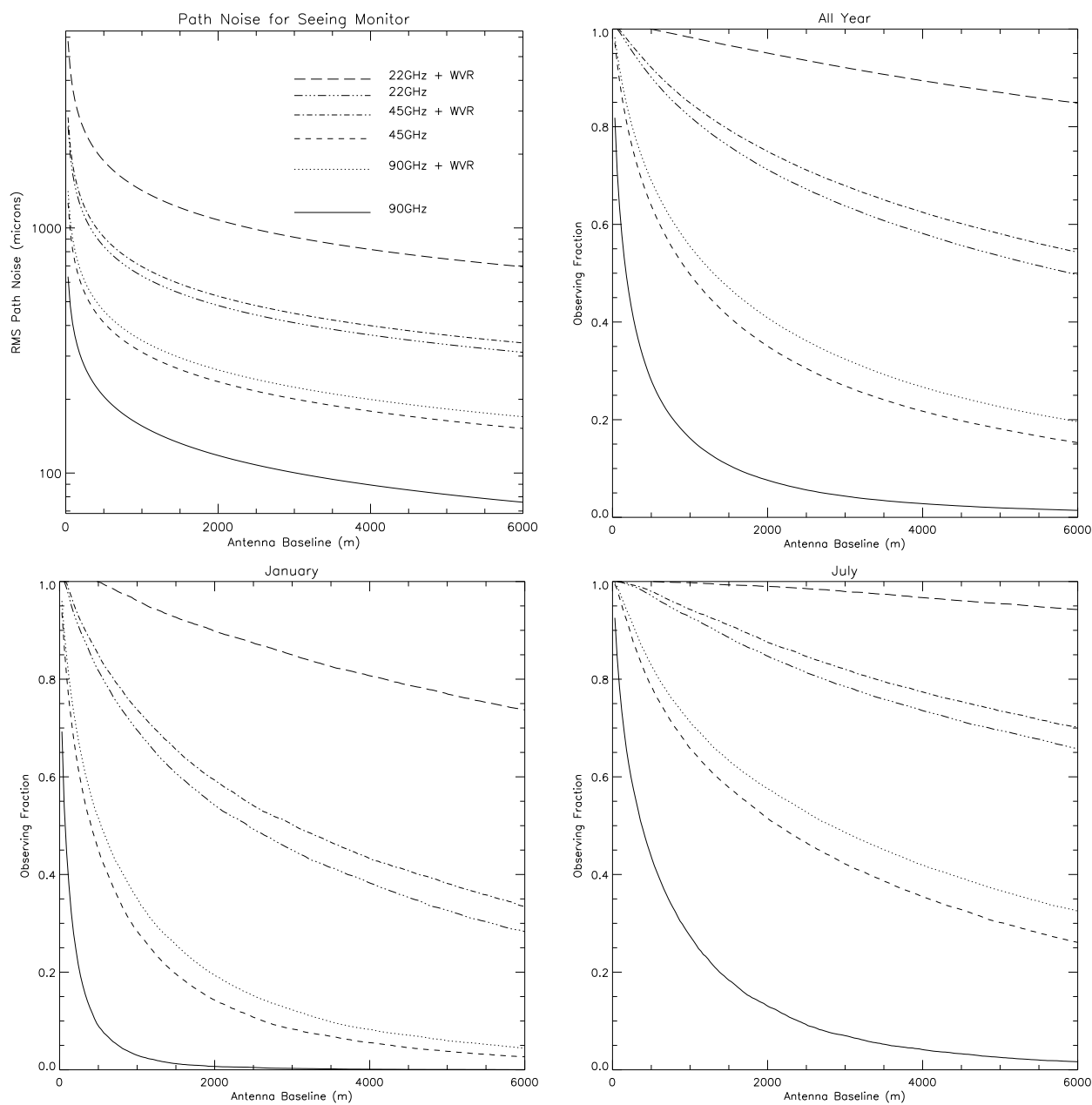}
\caption{Observing fractions as a function of antenna baseline, frequency and season for the ATCA, with and without the use of the water vapour radiometers (WVRs) to provide mm-wave phase correction. The top-left plot shows the rms path length noise as measured by the seeing monitor that is required to conduct successful observations on a given baseline and frequency, with and without the WVRs. The other three plots convert these into a fraction of the available observing time that could be used for the three cases: (i) the yearly average, (ii) January (i.e. summer) and (iii) July (i.e. winter). The calculations have been performed for 3 frequencies: 22, 45 \& 90 GHz and assume the median value for the Kolmogorov exponent $\alpha = 0.4$. As described in \S\ref{sec:efficiency}, this Figure can be used together with Table~\ref{table1} to estimate observing time fractions for any time band and month at one of these frequencies, on an antenna baseline of interest.}
\label{figure11}
\end{center}
\end{figure*}

To successfully undertake interferometry on a given baseline and frequency, the visibility efficiency 
\begin{equation}
\epsilon = \frac{<V_m>}{V} =  e^{-\sigma_{\phi}^2 / 2}, 
\label{eqn:viseff}
\end{equation}
where $\sigma_{\phi}$ is the phase fluctuation rms and $V$ the visibility, needs to be $\gtrsim 0.9$ (this is equivalent to a phase noise $\sigma_{\phi} = 30^{\circ}$).  From this value the limiting rms path noise fluctuation can be determined for a given waveband ($d_{rms} = \frac{\phi_{rms}}{360} \lambda_{obs}$).  In turn, applying Kolmogorov turbulence theory (i.e.\ \S\ref{sec:theory}) the maximum rms path noise on the 230\,m seeing monitor baseline can be determined for interferometry to be possible for a given frequency and baseline, applying eqn.~\ref{eqn:kolmogorov}, as:

\begin{equation}
\label{eqn:kolmogorov}
d_{rms} = d_{seemon-rms} \times \left(\frac{b}{b_{seemon}}\right)^{\alpha},
\end{equation}
where $b_{seemon} = 230$\,m.  We have taken $\alpha = 0.4$, the median value found for the Kolmogorov index. These limiting values for the path fluctuations are shown in the top-left plot of Fig.~\ref{figure11} for observations at 22, 45 and 90\,GHz.

Furthermore, a set of water vapour radiometers (WVRs) has recently been developed for the ATCA (see \citet{Indermuehle:2013fk}).  In that paper it is demonstrated that phase fluctuations which produce a visibility efficiency of 0.5 (i.e.\ phase fluctuations of $67^{\circ}$) can be corrected to provide a visibility efficiency of 0.9 using the WVRs.  This, in turn, can be used to determine the limiting rms path fluctuations between the seeing monitors for successful observations when the WVRs are in use.  These values are also shown in the top-left plot of Figure \ref{figure11}.  As can be seen, the limiting rms path values with WVRs in use are significantly higher than the limits without.

The rms path lengths we have presented in this paper can also be used to determine for what fraction of the time it is possible to observe on a given frequency and baseline as a function of hour and/or season, with and without the use of the WVRs.  The remaining three plots in Fig.\ \ref{figure11} show the results for three cases: (i) averaged over the whole year, (ii) in January and (iii) in July.

Some striking conclusions can be drawn from inspection of these plots. For instance, at 90\,GHz observing efficiency using WVRs becomes similar to that at 45\,GHz without their use.  At 90\,GHz observing efficiency on a 3\,km baseline is only possible for about 7\% of the time in winter, and so is only rarely attempted.  This could rise to half the time if WVRs were to be used.  45\,GHz interferometry is possible out to 6\,km baselines in winter for one-quarter of the time, and only possible in summer on short baselines of less than 1\,km.  With the use of WVRs, 45\,GHz observations in summer could become a regular practice out to 3 \,km baselines (and at 6\,km for one-third of the time).  22\,GHz interferometry can be conducted through the winter on all baselines for two-thirds of the time, but only for one quarter of the time in summer. With the use of WVRs it could be conducted on all baselines, for most of the time. 

Table~\ref{table1} and Figure~\ref{figure11} can also be used together to estimate the useable observable fractions in every 3 hour time band for each month of the year.  To do so, use the top-left plot in Figure~\ref{figure11} to determine the maximum rms path length variations for the observing frequency (i.e.\ 22, 45 or 90\,GHz) and antenna baseline of interest, both with and without the use of WVRs.  For the time band and month under consideration for an observing programme compare these values to the 25\%, 50\% and 75\% quartiles listed in Table~\ref{table1} to find the relevant observing fractions. For example, for a 3\,km antenna baseline at 90\,GHz, from Figure~\ref{figure11} we infer that path fluctuations of up to 100$\mu$m and 220$\mu$m can be used, without and with the use of WVRs, respectively. Without WVRs, from Table~\ref{table1} it can be seen that the lowest quartile value for June is 118$\mu$m, so 90 GHz interferometry on a 3 km baseline will only rarely be possible.  Table 1 also shows that the median night time path fluctuation are less than 200$\mu$m through the winter months. Thus in winter, with the use of WVRs interferometry at 90 GHz on this baseline will be possible during nights for more than 50\% of the time, and indeed in daylight for more than 25\% of the time (aside from between 12--15 hrs).  In January, on the other hand, such observations should only attempted during night time hours between 21--06 hrs, but even then will only be possible for less than 25\% of the time.

In summary, significant gains in observing time and increases in useable baseline length are seen to be achievable in all seasons by using WVRs for all three of the millimetre observing bands.  The useable observing period for a given baseline and frequency is increased by about 4 months a year.  This would make interferometry possible throughout the year in all the millimetre wavebands.  

\section{CONCLUSION}
The magnitude of the path length fluctuations caused by variations in the water vapour columns between two antennae of an interferometer determines when observations may be successfully attempted at a given frequency.  We have analysed 8.5 years of data from a seeing monitor at the ATCA site near Narrabri to provide statistics on the site conditions for this telescope, extending the initial 1 year study conducted by \citet{Middelberg:2006vn}. 

Two monitors located 230\,m apart measure the signal from a geosyncronous satellite.  We use the fluctuations in the path differences across this baseline to determine the rms value for the path difference, the Kolmogorov exponent, $\alpha$, for the turbulence and the phase screen speed, under the ``frozen-screen'' assumption \citep{Taylor:1938fk} for the passage of the turbulent cells across the site.  We also determine the seeing at 3.3\,mm from the half power points in the visibility.

Considerable variations in the rms path differences between the two monitors are found between summer and winter, as well as between day and night time, with 25\% (first) quartile values ranging from a low of 111$\mu$m in May between 03--06 hrs, to a high of 653$\mu$m in February between 12--15 hrs.  However, typical fluctuations during summer nights, $\sim 300 \mu$m, are similar to those occuring during winter days; in other words mm-wave interferometry can typically be undertaken during summer nights.  Variations in the Kolmogorov exponent are much less, with a median value of $\sim 0.4$ and variations of $\sim \pm 0.04$.  Similarly, variations in the phase screen speed are relatively small, with mode values found of $\sim 2$\,m/s.

The Kolmorogorov exponent is usually closer to the value expected for 2D turbulence, $1/3$, than that for 3D turbulence, $5/6$, consistent with the turbulent layer containing the fluctuating water vapour being relatively thin, as per the frozen-screen hypothesis.  The lowest values for $\alpha$ are found in winter, as would be expected for a less \textbf{agitated} troposphere than in summer.

The seeing at 3.3\,mm shows significant seasonal and time of day variations, depending on the path length rms values.  It ranges from a low of $\sim 0.3''$ during winter nights to $\sim 3''$ in the middle of summer days.

We have calculated the maximum path length fluctuations that can be tolerated for interferometric observations to be conducted with the ATCA on a given antenna baseline and frequency (in particular, the three receivers at 22, 45 and 90\,GHz).  Using the measured path length fluctuations, and assuming Kolmogorov turbulence with the median value measured for  $\alpha$ of 0.4, we  then estimate the useable observing fractions for time of day and month of the year.  We also do this assuming that phase variations can be partially rectified using the ATCA water vapour radiometers (WVRs).   Gains are possible on all baselines, with useable observing periods typically increased by about 4 months per year for any given combination of frequency/baseline using the WVRs.  In particular, observations at 90\,GHz could be conducted out to the currently maximum 3\,km baseline of the ATCA during winter months, or out to 6\,km were the six kilometre antenna to be equipped with 3mm capability.

With the use of WVRs, observations will often be undertaken with baselines longer than now being used. This would then facilitate flexible observing as when the conditions are not suitable for mm observations, the longer baselines will be conducive for many centimetre band programmes, as was proposed by 
\citet{Hall:1992fk} in their original proposal for operation of the ATCA at millimetre wavelengths.

\begin{acknowledgements}
The Australia Telescope Compact Array radio telescope is part of the Australia Telescope National Facility which is funded by the Commonwealth of Australia for operation as a National Facility managed by CSIRO. Development of the Water Vapour Radiometers (WVRs) on the ATCA were funded by the Australian Research Council LIEF grant funding scheme (project number LE0882778) with the support of UNSW, Sydney, Swinburne and JCU universities, as well as the ATNF.
\end{acknowledgements}

\section{TABLES AND FIGURES}
\clearpage

\clearpage

\clearpage

\clearpage

\clearpage

\clearpage

\clearpage

\clearpage

\clearpage

\clearpage

\clearpage

\begin{table}\scriptsize
\begin{tabular}{lcccccccccccc}
\toprule
/ Time & \multicolumn{8}{c}{25\%, 50\% and 75\% Quartiles in $\mu$m} \\
\cmidrule(r){2-9}
Month  & 00 & 03 & 06 & 09 & 12 & 15 & 18 & 21 & Min & Hour & Max & Hour \\
\midrule
     &      229 & 212 & 265 & 465 & 621 & 543 & 319 & 251 & 212 & 3 & 621 & 12 \\
January &   303 & 287 & 356 & 661 & 883 & 722 & 430 & 347 & 287 & 3 & 883 & 12 \\
    &       423 & 412 & 495 & 945 & 1243 & 1026 & 604 & 478 & 412 & 3 & 1243 & 12 \\
\midrule
  &        221 & 205 & 250 & 467 & 653 & 525 & 303 & 254 & 205 & 3 & 653 & 12 \\
February &   296 & 281 & 357 & 660 & 931 & 725 & 414 & 338 & 281 & 3 & 931 & 12 \\
  &         417 & 404 & 487 & 935 & 1259 & 985 & 552 & 477 & 404 & 3 & 1259 & 12 \\
\midrule
  &        195 & 177 & 200 & 407 & 582 & 427 & 253 & 215 & 177 & 3 & 582 & 12 \\
March  &     262 & 248 & 275 & 559 & 800 & 589 & 346 & 291 & 248 & 3 & 800 & 12 \\
 &          359 & 350 & 382 & 782 & 1053 & 822 & 467 & 422 & 350 & 3 & 1053 & 12 \\
\midrule 
 &        150 & 144 & 153 & 308 & 523 & 311 & 193 & 161 & 144 & 3 & 523 & 12 \\
April &          206 & 196 & 212 & 434 & 728 & 431 & 253 & 221 & 196 & 3 & 728 & 12 \\
 &          296 & 293 & 303 & 639 & 958 & 614 & 356 & 311 & 293 & 3 & 958 & 12 \\
\midrule
 &        116 & 111 & 112 & 197 & 370 & 219 & 145 & 125 & 111 & 3 & 370 & 12 \\
May &          159 & 154 & 160 & 305 & 528 & 306 & 194 & 171 & 154 & 3 & 528 & 12 \\
 &          224 & 234 & 233 & 476 & 752 & 438 & 269 & 236 & 224 & 0 & 752 & 12 \\
\midrule
 &        125 & 118 & 122 & 194 & 328 & 211 & 145 & 139 & 118 & 3 & 328 & 12 \\
June &          180 & 177 & 173 & 286 & 461 & 292 & 200 & 195 & 173 & 6 & 461 & 12 \\
 &          269 & 270 & 258 & 405 & 628 & 405 & 282 & 277 & 258 & 6 & 628 & 12 \\
\midrule
 &        121 & 114 & 118 & 198 & 370 & 226 & 149 & 133 & 114 & 3 & 370 & 12 \\
July &          171 & 162 & 162 & 294 & 497 & 317 & 208 & 178 & 162 & 6 & 497 & 12 \\
 &          250 & 237 & 251 & 433 & 671 & 441 & 288 & 251 & 237 & 3 & 671 & 12 \\
\midrule
 &        121 & 111 & 117 & 256 & 419 & 254 & 150 & 128 & 111 & 3 & 419 & 12 \\
August &          167 & 161 & 166 & 371 & 574 & 356 & 210 & 176 & 161 & 3 & 574 & 12 \\
 &          259 & 242 & 252 & 550 & 789 & 512 & 294 & 256 & 242 & 3 & 789 & 12 \\
\midrule
 &        130 & 126 & 148 & 371 & 482 & 288 & 167 & 143 & 126 & 3 & 482 & 12 \\
September &          191 & 184 & 228 & 532 & 679 & 415 & 234 & 202 & 184 & 3 & 679 & 12 \\
 &          289 & 282 & 348 & 757 & 945 & 571 & 335 & 284 & 282 & 3 & 945 & 12 \\
\midrule
 &        144 & 139 & 190 & 416 & 516 & 333 & 183 & 159 & 139 & 3 & 516 & 12 \\
October &          212 & 199 & 284 & 592 & 725 & 471 & 251 & 226 & 199 & 3 & 725 & 12 \\
 &          307 & 305 & 423 & 875 & 993 & 662 & 374 & 341 & 305 & 3 & 993 & 12 \\
\midrule
 &       204 & 191 & 268 & 478 & 583 & 398 & 239 & 214 & 191 & 3 & 583 & 12 \\
November &          273 & 263 & 363 & 678 & 808 & 557 & 336 & 298 & 263 & 3 & 808 & 12 \\
 &          410 & 362 & 507 & 947 & 1103 & 796 & 490 & 442 & 362 & 3 & 1103 & 12 \\
\midrule
 &        196 & 192 & 261 & 504 & 589 & 456 & 253 & 209 & 192 & 3 & 589 & 12 \\
December &          280 & 270 & 367 & 704 & 873 & 643 & 350 & 295 & 270 & 3 & 873 & 12 \\
 &          418 & 396 & 536 & 1008 & 1187 & 890 & 505 & 440 & 396 & 3 & 1187 & 12 \\
\bottomrule
\end{tabular}
\caption{Quartile values for the rms zenith path differences over the 230\,m seeing monitor baseline determined for the 8.5 year dataset from April 2005 to October 2013.  The data has been split by time of day into three hour intervals (the starting time for each is listed; i.e.\ $00 \equiv 0-3$\,hrs, etc.) and month. Also shown are the maximum and minimum values and the hours when they occurr, for each quartile and month combination. Times are in AEST.}
\label{table1}
\end{table}
\clearpage

\begin{table}
\begin{tabular}{lcccccccc}
\toprule
/ Time & \multicolumn{8}{c}{Minimum Quartiles in $\mu$m} \\
\cmidrule(r){2-9}
Quartiles  & 00 & 03 & 06 & 09 & 12 & 15 & 18 & 21 \\
\midrule

\multirow{2}{*}{25\%}  &  116  &  111  &  112  &  194  &  328  &  211  &  145  & 125 \\
      &    {\it May}   &    {\it May}  &     {\it May}    &   {\it Jun}   &     {\it Jun}e   &      {\it Jun}e   &     {\it May}   &    {\it May}\\
\multirow{2}{*}{50\%}   & 159  &  154  &  160   & 286  &  461  &  292  &  194  &  171 \\
 &        {\it May}& {\it May}&       {\it May} &      {\it Jun}&      {\it Jun}&      {\it Jun}&      {\it May}&      {\it May}\\
\multirow{2}{*}{75\%} &  224  &  234  &  233  &  405  &  628  &  405  &  269  &  236 \\
 &        {\it May}&      {\it May}&      {\it May}&      {\it Jun}&      {\it Jun}&      {\it Jun}&      {\it May}&      {\it May}\\
\midrule
 & \multicolumn{8}{c}{Maximum Quartiles in $\mu$m} \\ 
 \cmidrule(r){2-9}
\multirow{2}{*}{25\%} &  229  &  212  &  268   & 504  &  653 &   543  &  319  &  254 \\
 &        {\it Jan}&      {\it Jan}&      {\it Nov}&      {\it Dec}    &   {\it Feb}   &   {\it Jan}   &   {\it Jan}   &   {\it Feb}\\
\multirow{2}{*}{50\%}  &  303  &  286  &  367  &  704  &  931  &  725 &   430  &  347 \\
 &        {\it Jan}   &   {\it Jan}   &   {\it Dec}  &      {\it Dec} &      {\it Feb}&      {\it Feb}&      {\it Jan}&      {\it Jan}\\
\multirow{2}{*}{75\%} &  423  &  412   & 536  & 1008  & 1259  & 1026  &  604  &  478 \\
 &        {\it Jan}&      {\it Jan}&      {\it Dec}    &  {\it Dec}   &   {\it Feb}   &   {\it Jan}   &   {\it Jan}   &   {\it Jan}\\

\bottomrule
\end{tabular}
\caption{The maximum and minimum quartiles for each 3 hour time band, together with the month in which they occur, as extracted from Table \ref{table1}.}
\label{table2}
\end{table}
\clearpage

\begin{table}
\begin{tabular}{lccccccccc}
\toprule
 & \multicolumn{8}{c}{Kolmogorov exponent $\alpha$} \\
\cmidrule(r){2-10}
Month / Hour band  & 00-03 & 03-06 & 06-09 & 09-12 & 12-15 & 15-18 & 18-21 & 21-24 & Median \\
\midrule
January   &   0.388  &   0.385  &   0.397   &  0.449   &  0.466  &   0.449  &   0.422  &   0.400  & 0.419\\
July   &   0.385  &   0.384  &   0.383   &  0.412   &  0.446   &  0.424   &  0.395  &   0.387  & 0.402\\
\midrule
Overall &  0.387   &  0.384   &  0.389  &   0.438  &   0.458   &  0.441   &  0.410  &   0.396  &   0.413\\
\bottomrule
\end{tabular}
\caption{The median Kolmogorov exponents $\alpha$ for each 3 hour time band during January and July, and over the whole year.}
\label{table4}
\end{table}
\clearpage

\begin{table}
\begin{tabular}{lcccc}
\toprule \\
& \multicolumn{2}{c}{Seeing $\theta$ [$''$]} \\
\cmidrule(r){2-3} \\
& \multicolumn{1}{c}{\it{June}} & \multicolumn{1}{c}{\it{December}} \\
Time AEST [h] & $\overline{\theta}$ & $\overline{\theta}$ \\
\midrule
 0-- 1 & 0.32 & 0.53\\
 1-- 2 & 0.33 & 0.50\\
 2-- 3 & 0.32 & 0.48\\
 3-- 4 & 0.33 & 0.49\\
 4-- 5 & 0.34 & 0.52\\
 5-- 6 & 0.35 & 0.59\\
 6-- 7 & 0.35 & 0.80\\
 7-- 8 & 0.36 & 1.18\\
 8-- 9 & 0.44 & 1.68\\
 9--10 & 0.65 & 2.13\\
10--11 & 0.91 & 2.48\\
11--12 & 1.12 & 2.56\\
12--13 & 1.19 & 2.53\\
13--14 & 1.10 & 2.27\\
14--15 & 0.90 & 2.04\\
15--16 & 0.65 & 1.69\\
16--17 & 0.49 & 1.36\\
17--18 & 0.42 & 1.03\\
18--19 & 0.38 & 0.78\\
19--20 & 0.38 & 0.69\\
20--21 & 0.36 & 0.64\\
21--22 & 0.34 & 0.64\\
22--23 & 0.33 & 0.66\\
23--24 & 0.31 & 0.65\\
\bottomrule
\end{tabular}
\caption{The hourly seeing values in arcseconds for June and December for observations at $\lambda$=3.3 mm. These are calculated using the median values determined for the rms path difference and Kolmogorov exponent for each of these time periods, as explained in \S\ref{sec:seeing}.}
\label{table3}
\end{table}
\clearpage


\bibliographystyle{apj}
\bibliography{WVR_REFS}



\end{document}